\documentclass[a4paper]{spie}  

\usepackage[]{graphicx}
\usepackage{subfigure}

\title{Parrondo's games with chaotic switching}

\author{T.W. Tang,\supit{a} A. Allison\supit{a} and D. Abbott\supit{a}
\skiplinehalf \supit{a} Center for Biomedical Engineering (CBME)
and Department of Electrical \& Electronic Engineering, The
University of Adelaide, SA 5005, Australia }

\authorinfo{Further author information: (Send correspondence to T.W.T.)\\T.W.T.: E-mail: tzewei@eleceng.adelaide.edu.au, Telephone: +61 8303 6296\\  A.A.: E-mail: aallison@eleceng.adelaide.edu.au, Telephone: +61 8303 5283\\ D.A.: E-mail: dabbott@eleceng.adelaide.edu.au, Telephone: +61 8303 5748}

\begin{document}
  \maketitle

\begin{abstract}
This paper investigates the different effects of chaotic switching
on Parrondo's games, as compared to random and periodic switching.
The rate of winning of Parrondo's games with chaotic switching
depends on coefficient(s) defining the chaotic generator, initial
conditions of the chaotic sequence and the proportion of Game A
played. Maximum rate of winning can be obtained with all the above
mentioned factors properly set, and this occurs when chaotic
switching approaches periodic behavior.
\end{abstract}



\section{INTRODUCTION}
\label{sect:intro}  

\subsection{Parrondo's games}\label{sect:games}
Parrondo's games were devised by the Spanish physicist Juan M. R.
Parrondo in 1996 and they were presented in unpublished form at a
workshop in Torino, Italy.\cite{Parrondo96} After about three
years, in 1999, Harmer and Abbott published the seminal paper on
Parrondo's games.\cite{Harmer99a} The games are named after their
creator and the counterintuitive behavior is called ``Parrondo's
paradox.".\cite{Harmer99b}

The main idea of Parrondo's paradox is that two individually
losing games can be combined to win via deterministic or
non-deterministic mixing of the games\cite{Harmer01}. There has
been a lot of research on Parrondo's games after the first
published paper, giving birth to new games such as history
dependent games\cite{Parrondo00} (instead of capital dependent)
and cooperative games\cite{Toral01} (multi-player games instead of
one player). However, in this paper the original Parrondo's games
will be used for analyzing the differences between chaotic, random
and periodic switching. The seminal paper that considered chaotic
switching in Parrondo's games was by Arena et al.\cite{Arena03}
The original Parrondo's games are defined as
below,\cite{Harmer99b,Harmer01,Harmer02,Harmer04} where $C$ is the
current capital at discrete-time step $n$.

\begin{description}
\item[Game A] consists of a biased coin that has a probability $p$
of winning,

\item[Game B] consists of 2 games, the condition of choosing
either one of the games is given as below:

    \begin{description}
      \item[$C$ mod $M = 0$] play a biased
      coin that has probability $p_1$ of winning,
      \item[$C$ mod $M \ne 0$]play a
      biased coin that has probability $p_2$ of winning.
     \end{description}

\end{description}
For the original Parrondo's games, the parameters are set as
below:
\[ M = 3,\]
\[ p = 1/2 - \epsilon,\]
\[ p_1 = 1/10 -\epsilon,\]
\[ p_2 = 3/4 -\epsilon.\]
To control the three probabilities $p$, $p_1$ and $p_2$, a biasing
parameter, $\epsilon$ is included in the above equations, where in
this paper $\epsilon$ is chosen to be 0.005.

\subsection{Chaos}\label{sect:chaos}
\begin{figure}[p]

  \subfigure[Logistic sequence with $a=4$.]{
  \label{fig:time:logis}
  \includegraphics[width=7cm]{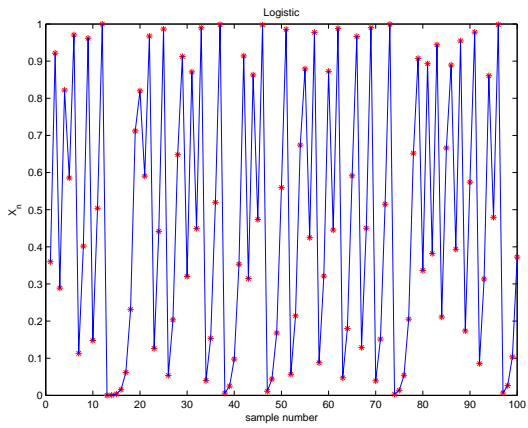}
  }
\hfill \subfigure[Random sequence.]{
  \label{fig:time:random}
  \includegraphics[width=7cm]{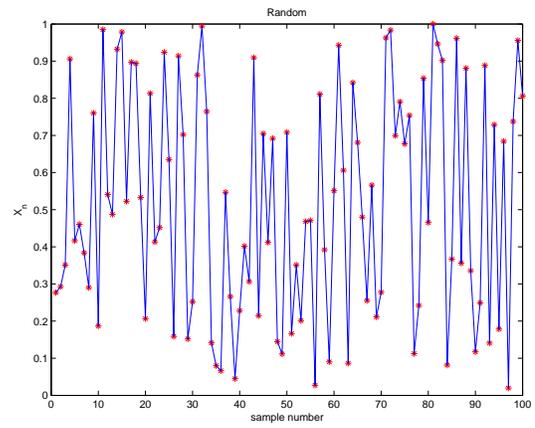}
  }
  \caption{Random and Logistic sequences.}\label{fig:time}
\end{figure}

\begin{figure}[p]

\subfigure[Phase space plot of Logistic sequence with $a=4$.]{
\label{fig:phase:logis}
  \includegraphics[width=7cm]{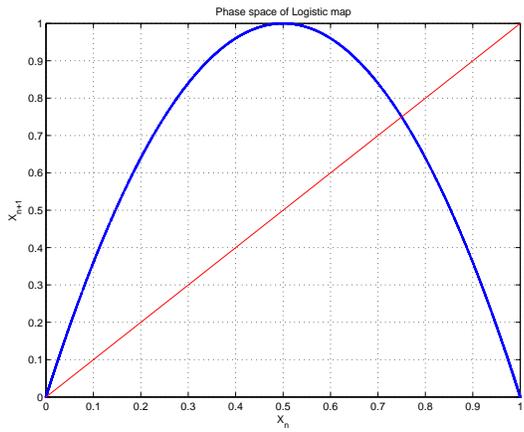}
  }
\hfill \subfigure[Phase space plot of random sequence.]{
\label{fig:phase:random}
  \includegraphics[width=7cm]{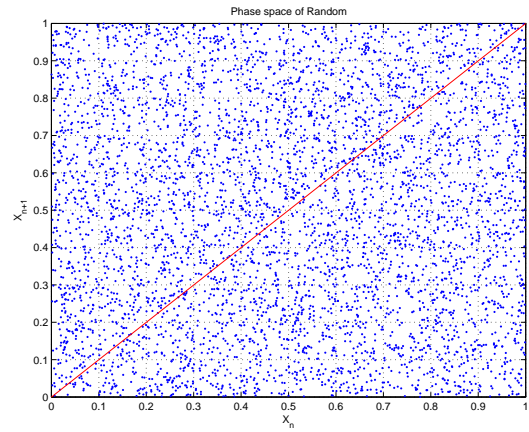}
  }
  \caption{Phase space plot of random and Logistic
sequences.}\label{fig:phase}
 \end{figure}

Chaos is used to describe fundamental disorder generated by simple
deterministic systems with only a few elements.\cite{Peitgen92}
The irregularities of chaotic and random sequences in the time
domain are often quite similar. As an illustration, the Logistic
sequence and random sequence are plotted as shown in Figure
\ref{fig:time:logis} and Figure \ref{fig:time:random} where it is
difficult to observe any difference.\footnote{The settings of all
parameters used in the MATLAB simulations for all figures in this
paper are reported in Table \ref{tab:figures}.} However, by
plotting the phase space of chaotic and random sequences, as shown
in Figure \ref{fig:phase:logis} and Figure \ref{fig:phase:random},
the chaotic sequence can be easily identified because there is a
regular pattern in the phase space plot. Consecutive points of a
chaotic sequence are highly correlated, but not for the case of a
pure random sequence. A chaotic sequence, $X$ is usually generated
by nested iteration of some functions. This is shown as below,
where $n$ is sample number and $f_n(\cdot)$ is $n^{th}$ iteration
of function $f(\cdot)$,
\begin{equation}
x_n=f_n(f_{n-1}(...f_2(f_1(x_0))...)).
\end{equation}

For simplicity, one-dimensional and two-dimensional chaotic maps
are used. More analysis on the Logistic Map is carried out in this
paper because it is one of the oldest and typical chaotic maps.
\begin{enumerate}
\item One-dimensional chaotic generators:
    \begin{description}
    \item[Logistic Map\cite{Parker89}] \begin{equation}\label{eq:logis} x_{n+1}= ax_n(1-x_n) \end{equation}
    \item[Tent Map\cite{Peitgen92}] \begin{equation}\label{eq:tent} x_{n+1}=\left\{\begin{array}{lll}
                                                                         ax_{n} & \mbox{if} \;x_n\le0.5\\
                                                                         a(1-x_{n}) &\mbox{otherwise}&
                                                                    \end{array}\right.
                    \end{equation}
     \item[Sinusoidal Map\cite{Arena03}] \begin{equation}\label{eq:sinus} x_{n+1}= ax_{n}^2\sin(\pi x_n) \end{equation}
     \item[Gaussian Map\cite{Bucolo02}] \begin{equation}\label{eq:gaus} x_{n+1}=\left\{\begin{array}{cll}
                                                                         0 & \mbox{if} & x_n=0\\
                                                                         \frac{1}{x_{n}}\;\mbox{mod}\;1 & \mbox{if} & x_n\ne 0
                                                                    \end{array}\right.
                    \end{equation}
    \end{description}

\item Two-dimensional chaotic generators:
    \begin{description}
    \item[Henon Map\cite{Peitgen92}] \begin{equation}\label{eq:henon} \left\{\begin{array}{ll}
                                                                    x_{n+1}=y_n+1-a{x_n}^2\\
                                                                    y_{n+1}=bx_n
                                                                    \end{array}\right.
                    \end{equation}

    \item[Lozi Map\cite{Bucolo02}] \begin{equation}\label{eq:lozi} \left\{\begin{array}{ll}
                                                                    x_{n+1}=y_n+1-a|x_n|\\
                                                                    y_{n+1}=bx_n
                                                                    \end{array}\right.
                    \end{equation}
    \end{description}
\end{enumerate}
\subsection{Switching strategies}
Parrondo's games consist of two games, Game A and Game B. The
definitions and rules of each of the games are explained in
Section \ref{sect:games}. At discrete-time step $n$, only one game
will be played, either Game A or Game B. The algorithm or pattern
used to decide which game to play at discrete-time step $n$ is
defined as the switching strategy. In the original Parrondo's
games, the switching strategies carried out are random and
periodic switchings.\cite{Harmer99b} In this paper, chaotic
switching of Game A and Game B based on several chaotic sequences
is investigated through simulations.

\section{Games with chaotic switching}
To play Parrondo's games with chaotic switching, a chosen chaotic
generator will be used to generate a sequence, $X$. Sequence $X$
will be used to decide either Game A or B to be played at
discrete-time step $n$. There are many ways to carry out this
task, but the easiest way is to compare each value of $X$ with a
constant $\gamma$. On each round (round $n$) of Parrondo's games,
a value from the chaotic sequence, $x_n$ is compared with
$\gamma$, if $x_n \le \gamma$, Game A will be played but if $x_n >
\gamma$, Game B will be played. This simple procedure is adopted
in this paper. For random switching, $\gamma$ is equivalent to the
proportion of Game A played after $n$ discrete-time steps.
However, for chaotic switching, $\gamma$ is not the proportion of
Game A played after discrete-time step $n$ unless the chaotic
sequence is uniformly distributed.

\subsection{Chaotic generators}
The outcomes of Parrondo's games will be affected by the different
chaotic switchings applied. Before this aspect is investigated,
the parameters that affect the behaviors of the chaotic sequence
have to be identified. The properties of a particular chaotic
sequence from a chaotic generator depend on two elements:
coefficient(s) of the chaotic generator and initial conditions.

\subsubsection{Coefficient(s) of a chaotic generator}\label{sect:vars}
\begin{figure}[ht]

 \subfigure[Bifurcation diagram of Logistic Map.]{
  \includegraphics[width=5cm]{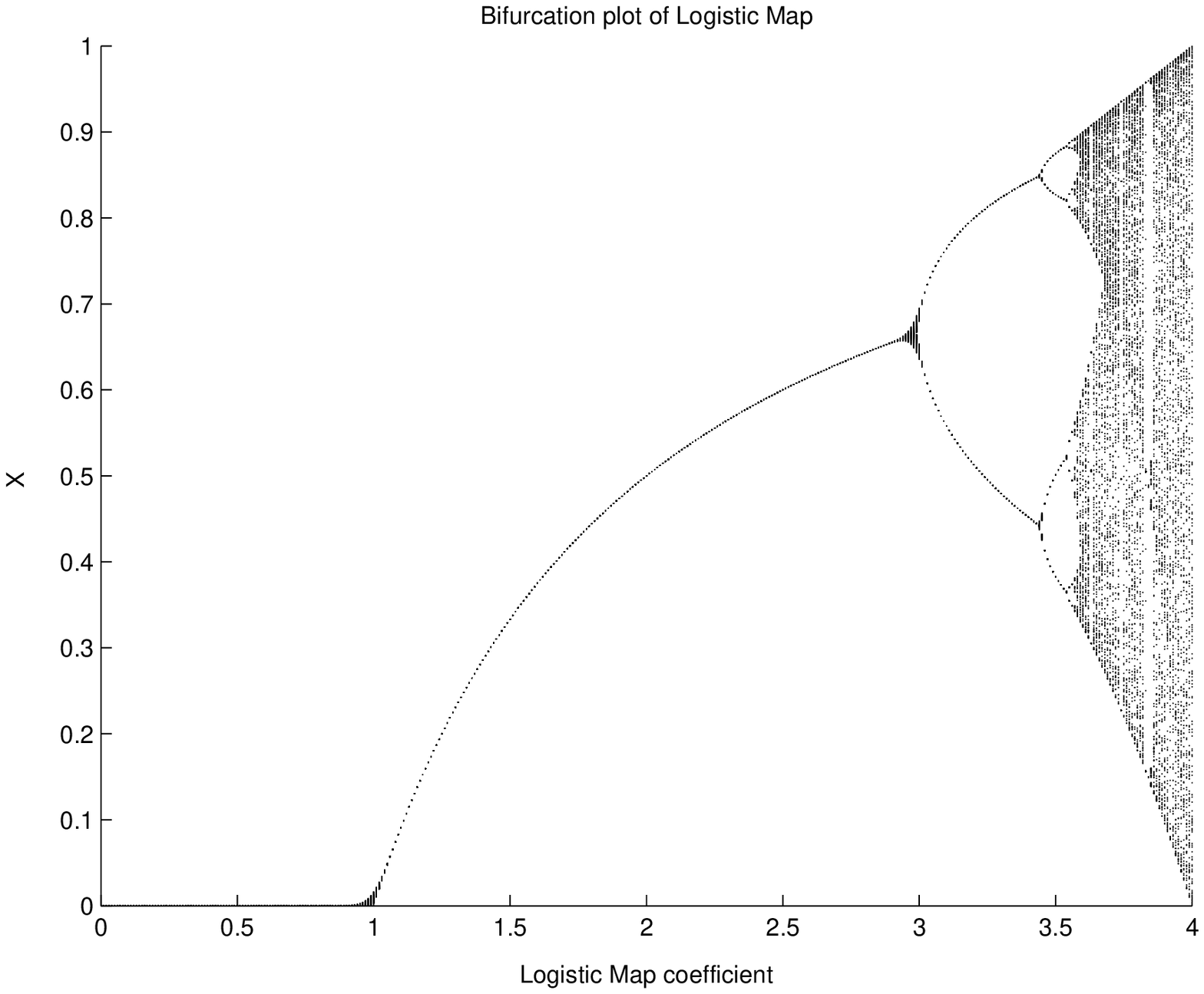}\\
  \label{fig:bifur:logis}}
\hfill \subfigure[Bifurcation diagram of Sinusoidal Map.]{
  \includegraphics[width=5cm]{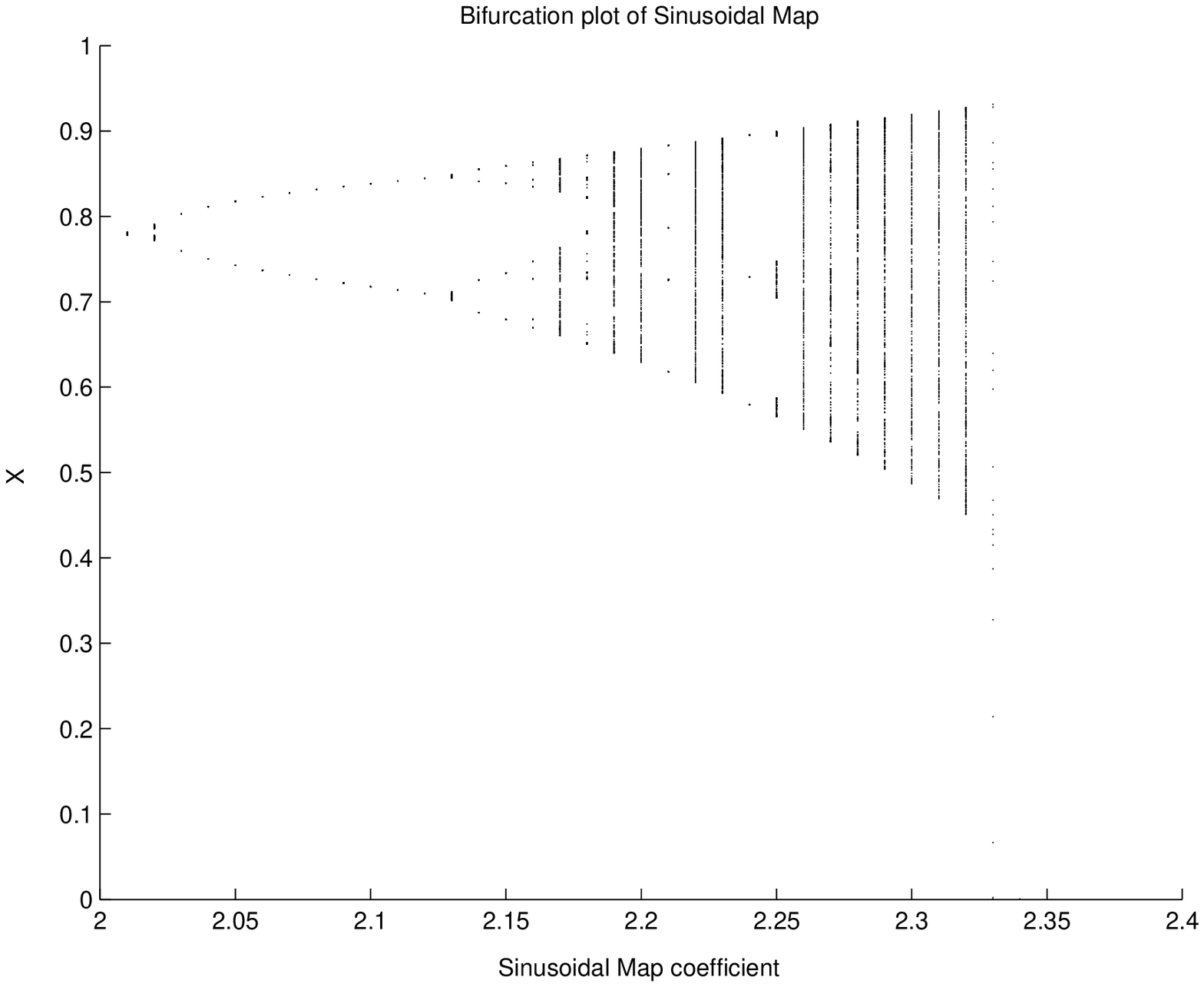}\\
  \label{fig:bifur:sinus}}
  \hfill \subfigure[Bifurcation diagram of Tent Map.]{
  \includegraphics[width=5cm]{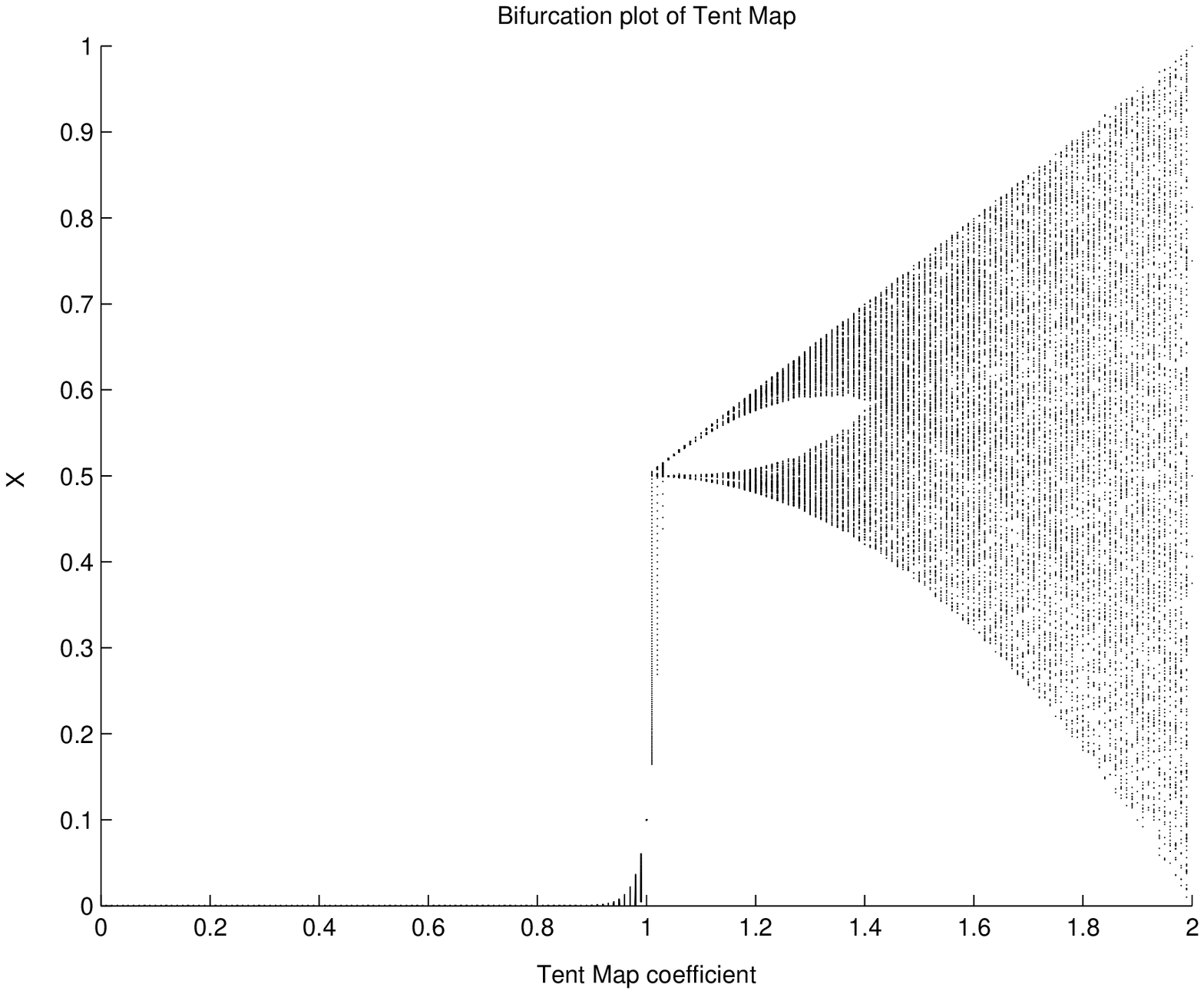}\\
  \label{fig:bifur:tent}}
\caption{Bifurcation diagrams of chaotic maps.}\label{fig:bifur}
\end{figure}

\begin{figure}[h]\begin{center}
  \includegraphics[width=10cm]{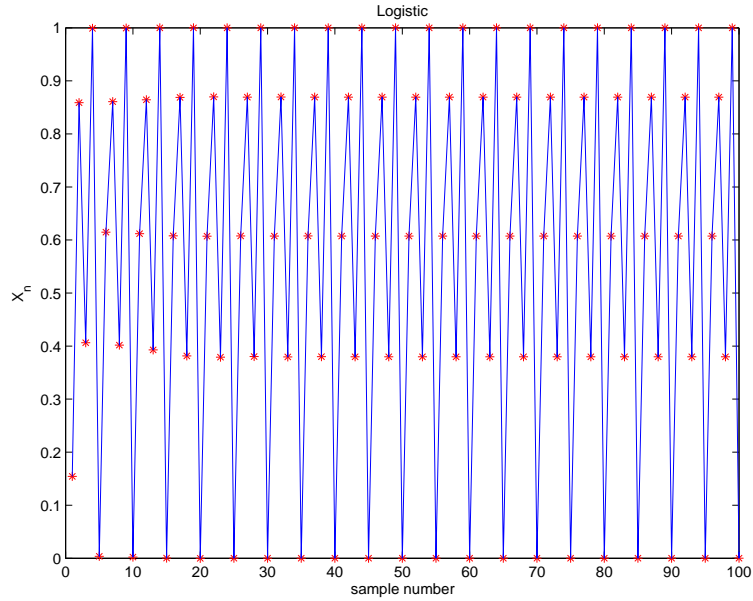}\\
  \caption{Periodic sequence from Logistic Map with $a=3.74$.}\label{fig:time:logis374}\end{center}
\end{figure}

The phase space of a chaotic signal changes with the
coefficient(s) defining its chaotic map as in Eq~(\ref{eq:logis})
to Eq~(\ref{eq:lozi}). That is the relationship between
consecutive points of a chaotic sequence changes with the
coefficient(s) and this can lead the chaotic sequence to either a
chaotic or stable state.\cite{Parker89} For example, the $a$
coefficient in a Logistic Map will decide the state of the
sequence generated, whether in a stable or chaotic state. This
result can be looked up from the bifurcation diagram of the
Logistic map as shown in Figure \ref{fig:bifur:logis}. The regions
with continuous points correspond to chaotic states, while those
with distinct points correspond to stable states.\cite{Parker89} A
Logistic sequence with $a=3.74$ is plotted in Figure
\ref{fig:time:logis374} to show the periodic behavior of the
sequence. The bifurcation diagrams of the one-dimensional chaotic
maps can be easily plotted as shown in Figure
\ref{fig:bifur:logis}, Figure \ref{fig:bifur:sinus} and Figure
\ref{fig:bifur:tent}. However, the complete bifurcation diagram of
two-dimensional chaotic maps are more complicated since 4
parameters are involved.


\begin{figure}[ht]

 \subfigure[Distribution of Logistic sequence in stable state ($a=3.74$).]{
  \includegraphics[width=7cm]{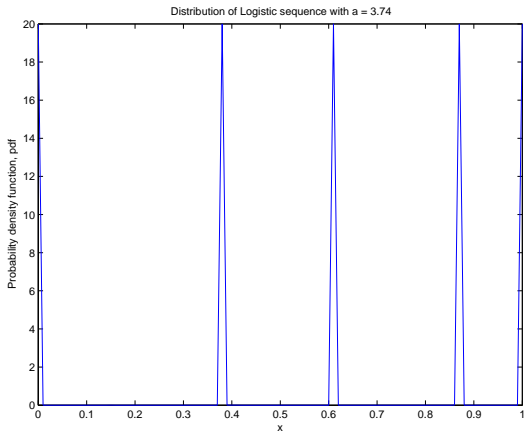}\\
  \label{fig:dist:logis374}}
\hfill \subfigure[Distribution of Logistic sequence in unstable
state ($a=4$).]{
  \includegraphics[width=7cm]{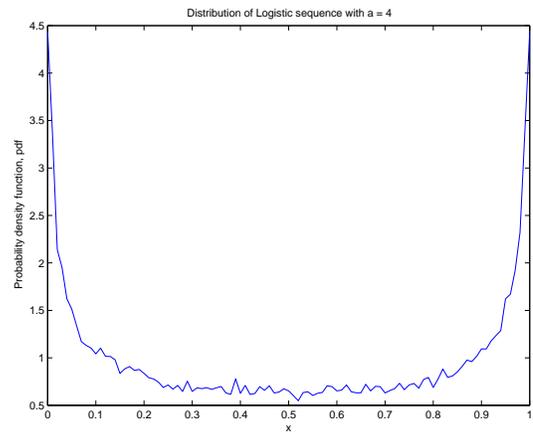}\\
  \label{fig:dist:logis}}
\caption{Distribution of Logistic sequence with different $a$
values.}\label{fig:dist}
\end{figure}

The distribution of $x_n$ of a chaotic sequence between 0 and 1 is
determined by the coefficients $a$ and $b$ as defined in
Eq~(\ref{eq:logis}) to Eq~(\ref{eq:lozi}). With coefficients that
correspond to stable states of a chaotic generator, only countable
distinct values of $x$ are observed. The pdf of $x_n$ for a
Logistic sequence in a stable state is plotted as shown in Figure
\ref{fig:dist:logis374}. In a chaotic state, the spread of $x_n$
is more even over the range from 0 to 1. This is shown with a
Logistic sequence and $a=4$, in Figure \ref{fig:dist:logis}.

\subsubsection{Initial conditions}\label{sect:initial}
Different initial conditions will give the same phase space plot
of the chaotic maps. However, initial conditions affect the way
the phase space is constructed. A small fluctuation in the initial
condition will start a ``snow ball" effect on the chaotic
sequence, which affects the values of the whole sequence after a
few iterations. This is one of the famous properties of a chaotic
sequence.\cite{Peitgen92}

There are some initial conditions that will map the chaotic
sequence to a constant value. An example to illustrate this
behavior is to use Logistic Map with $a=4$. From simulation
results, initial conditions of 0, 1/4, 1/2 and 3/4 will map the
sequence to a constant value of either 0 or 3/4. This can be
explained from the phase space plot of the Logistic Map as seen in
Figure \ref{fig:phase:logis}. The intersections of the phase space
plot and the line $x_{n+1}=x_n$ occur at $x_{n+1}=x_n=0$ and
$x_{n+1}=x_n=3/4$. These intersection points are the attractors of
Logistic Map with $a=4$. The choice of initial conditions in this
case drives the sequence directly to the attractors. Hence the
choice of initial conditions is vital for obtaining an oscillating
sequence in order to carry out effective chaotic switching of
Parrondo's games. Similar attractors can be observed using other
chaotic maps.\cite{Peitgen92}

\subsection{$\gamma$ values}
To decide whether to play Game A or B, at each discrete-time step
$n$, we utilize the $\gamma$ parameter. The $\gamma$ value sets a
threshold on selection of games to be played on each round. On the
other hand, $\gamma$ value is important to make Parrondo's paradox
appear. The effect of the $\gamma$ value on the rate of winning of
Parrondo's games with chaotic switching is investigated through
simulations. All the chaotic sequences are normalized to have
values between 0 and 1 for ease of comparison between chaotic
switchings with the same $\gamma$ value.

\section{Effect of the coefficient(s) of Chaotic generator on the Rate of
winning}\label{sect:win:vars} The rate of winning, $R(n)$, is
given below\cite{Harmer01}, where $\pi_j(n)$ is the stationary
probability of being in state $j$ at discrete-time step $n$.
\begin{equation}
R(n)=E[J_{n+1}-J_n]=E[J_{n+1}]-E[J_n]=\sum_{j=-\infty}^\infty
j[\pi_j(n+1)-\pi_j(n)].
\end{equation}

The coefficient(s) of a chaotic generator will determine the
stability of the chaotic system. Under the stable regions, the
system will show periodic behaviors. The simulation results show
that the maximum rate of winning of Parrondo's games occurs when
the chaotic generator used for switching tends toward periodic
behavior. On the other hand, when a chaotic generator behaves
truly chaotically, the rate of winning is smaller compared to a
periodic case. Hence, under periodic or stable state of a chaotic
sequence, and properly tuned initial conditions and $\gamma$ value
as discussed in the next section, the rate of winning obtained can
be higher than the one achieved by random switching. However, to
identify the exact periodic sequence that gives the highest rate
of winning is a complicated problem.

\begin{figure}[h]

 \subfigure[Gain of Henon switching after 100 games with different $a$ \& $b$ coefficients.]{
  \includegraphics[width=8cm]{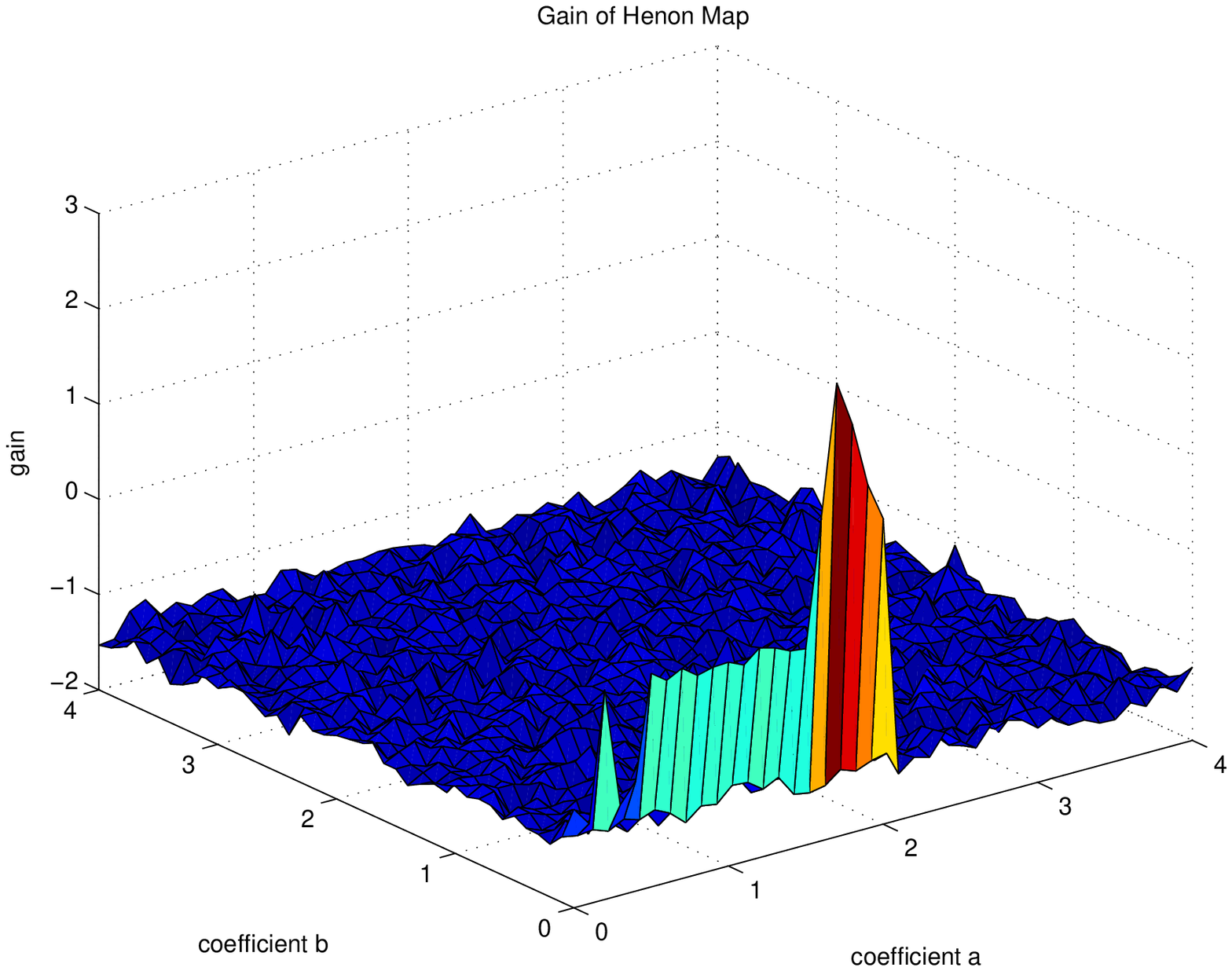}\\
  \label{fig:threeAB:henonAB}}
\hfill \subfigure[Gain of Lozi switching after 100 games with
different $a$ \& $b$ coefficients.]{
  \includegraphics[width=8cm]{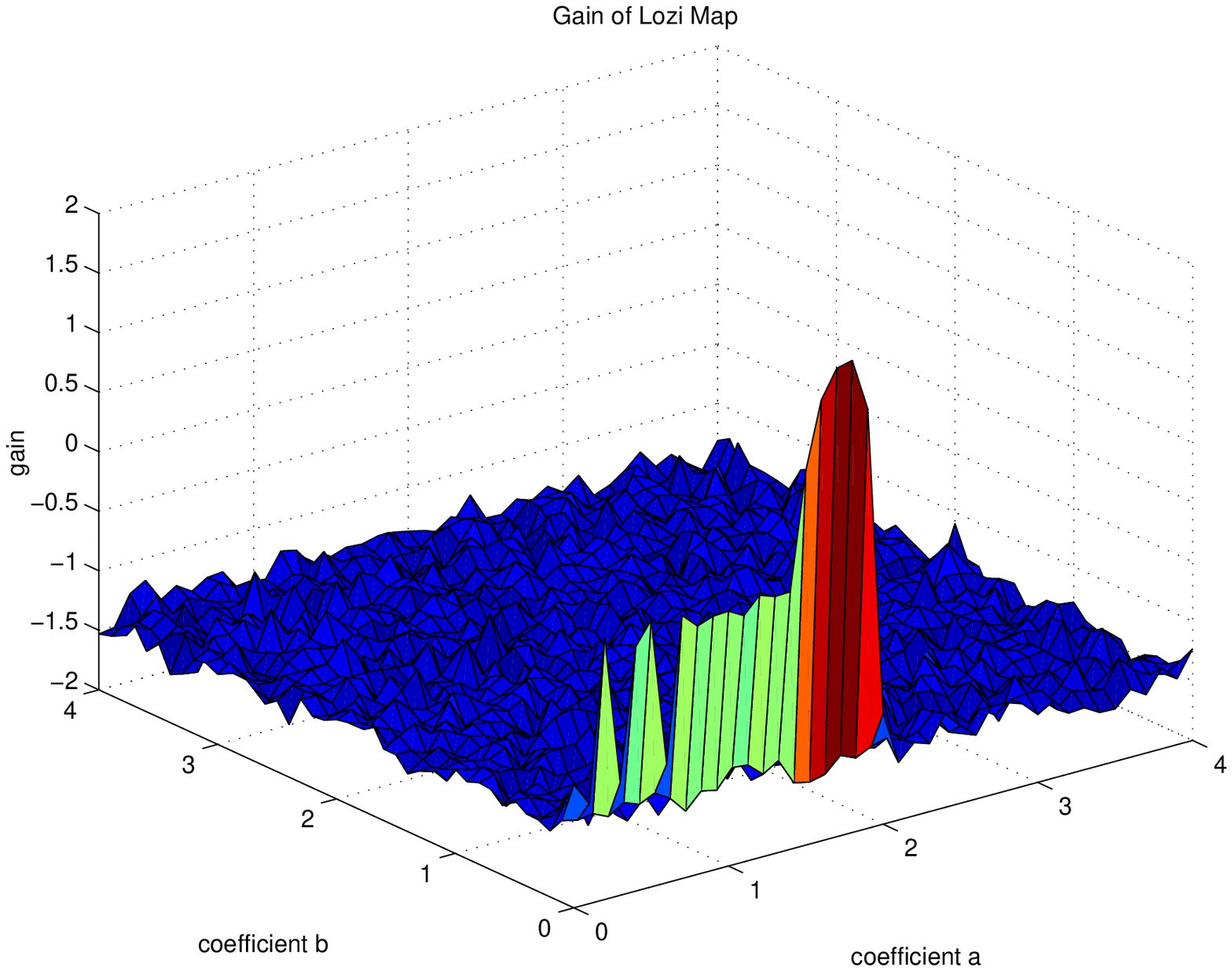}\\
  \label{fig:threeAB:loziAB}}
  \caption{Gain of Parrondo's games with switching based on
two-dimensional maps against $a$ and $b$
coefficients.}\label{fig:threeAB}
\end{figure}

For two-dimensional maps such as the Henon Map and Lozi Map, there
are two coefficients, $a$ and $b$ that control the behavior of the
sequence. Hence, they determine the rate of winning of Parrondo's
games. From Figure \ref{fig:threeAB:henonAB} and Figure
\ref{fig:threeAB:loziAB}, the gains after 100 games are plotted
with different combinations of $a$ and $b$ values. It is found
that both the maps give maximum gain after 100 games when $a=1.7$
and $b=0$. For $b=0$, Eq~\ref{eq:henon} and Eq~\ref{eq:lozi} are
simplified to Eq~\ref{eq:henon1} and Eq~\ref{eq:lozi1}
respectively, which are one-dimensional.
\begin{equation}\label{eq:henon1}
    x_{n+1}=1-a{x_n}^2,\\
    \end{equation}
\begin{equation}\label{eq:lozi1}
    x_{n+1}=1-a|x_n|.\\
    \end{equation}

\section{Effect of initial conditions and $\gamma$ value on the rate of winning}
\begin{figure}[ht]

 \subfigure[Gain of Logistic switching after 100 games.]{
  \includegraphics[width=6cm]{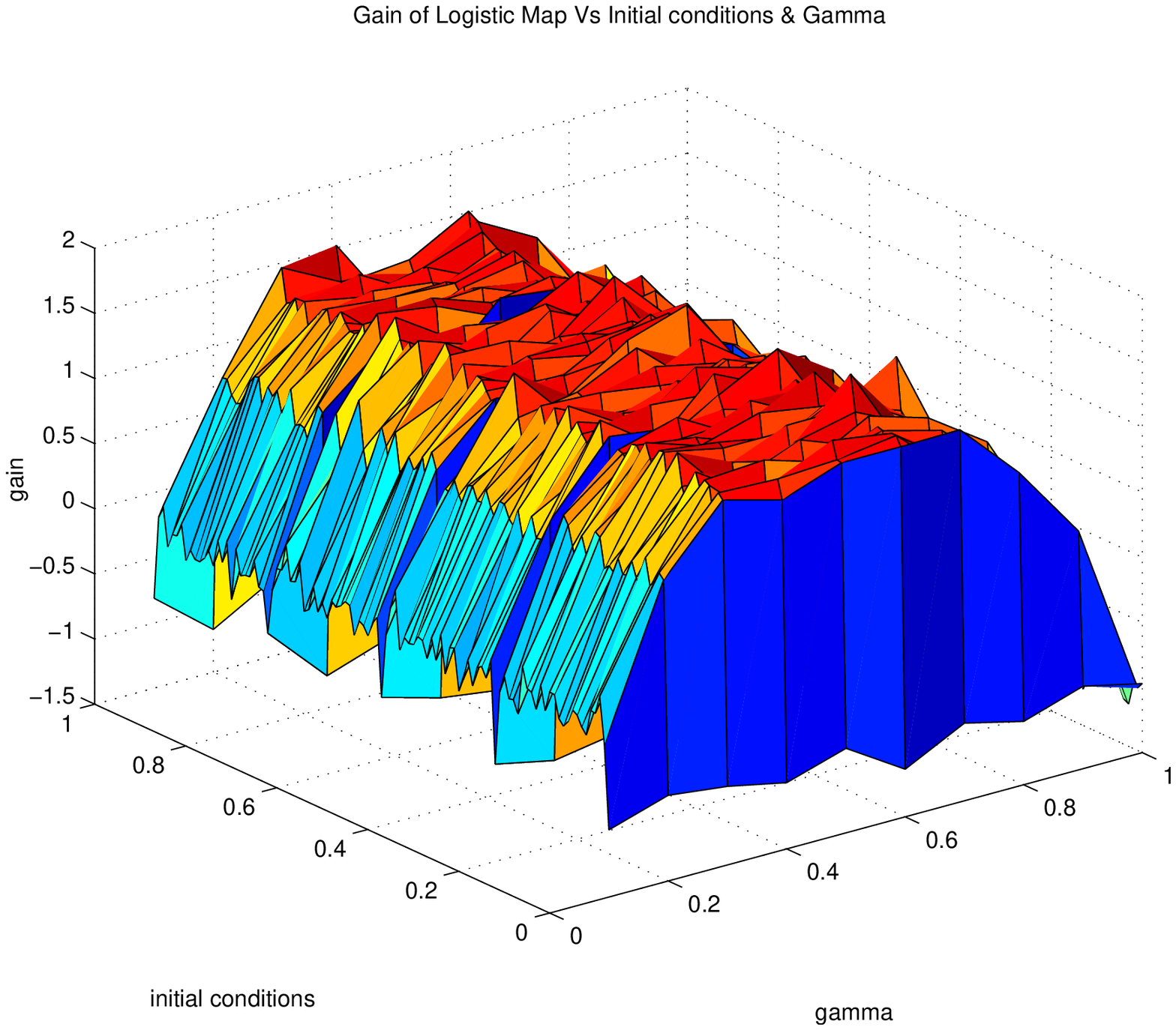}\\
  \label{fig:three:logis}}
\hfill \subfigure[Gain of Sinusoidal Map switching after 100
games.]{
  \includegraphics[width=6cm]{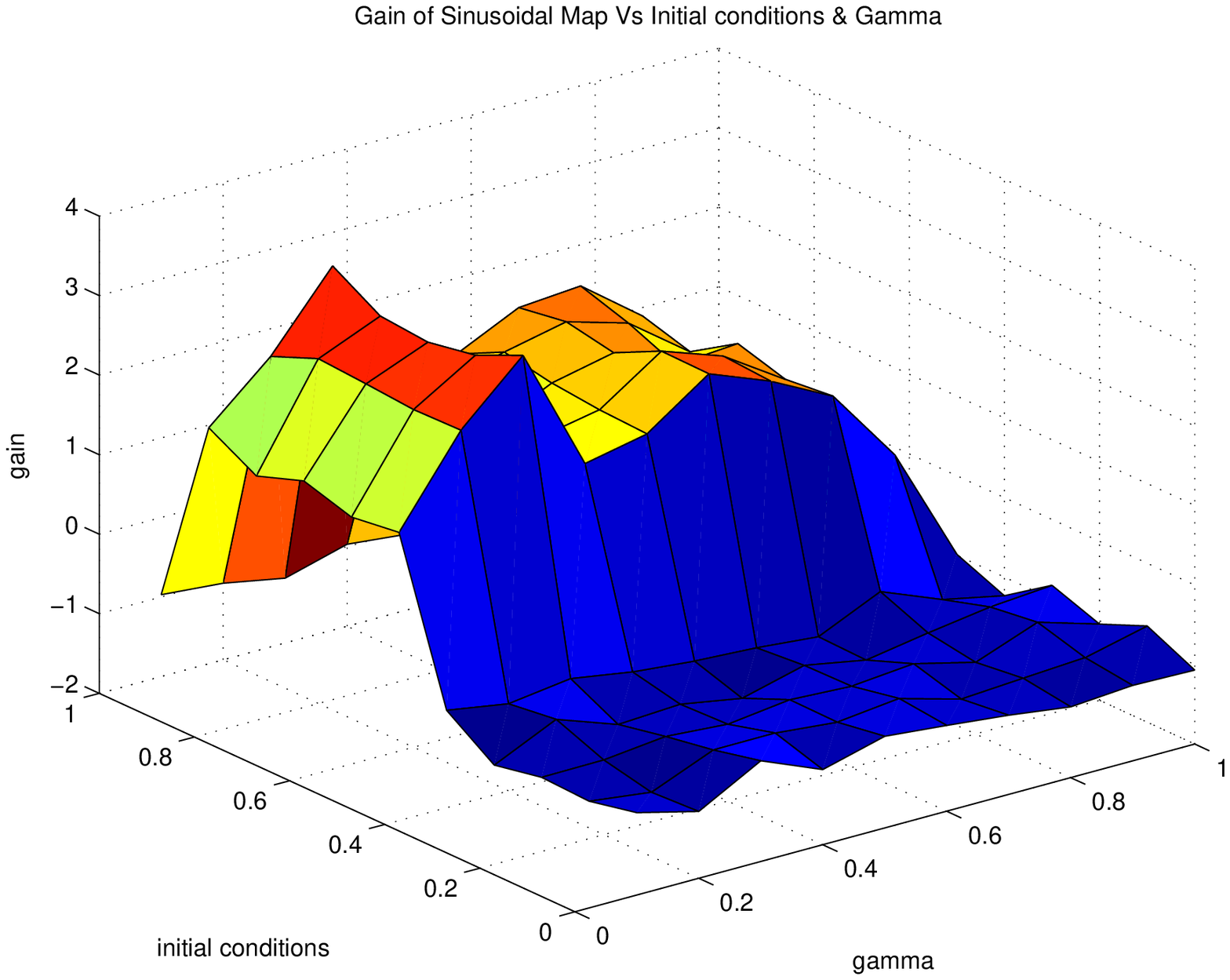}\\
  \label{fig:three:sinus}}
\hfill \subfigure[Gain of Tent Map switching after 100 games.]{
  \includegraphics[width=6cm]{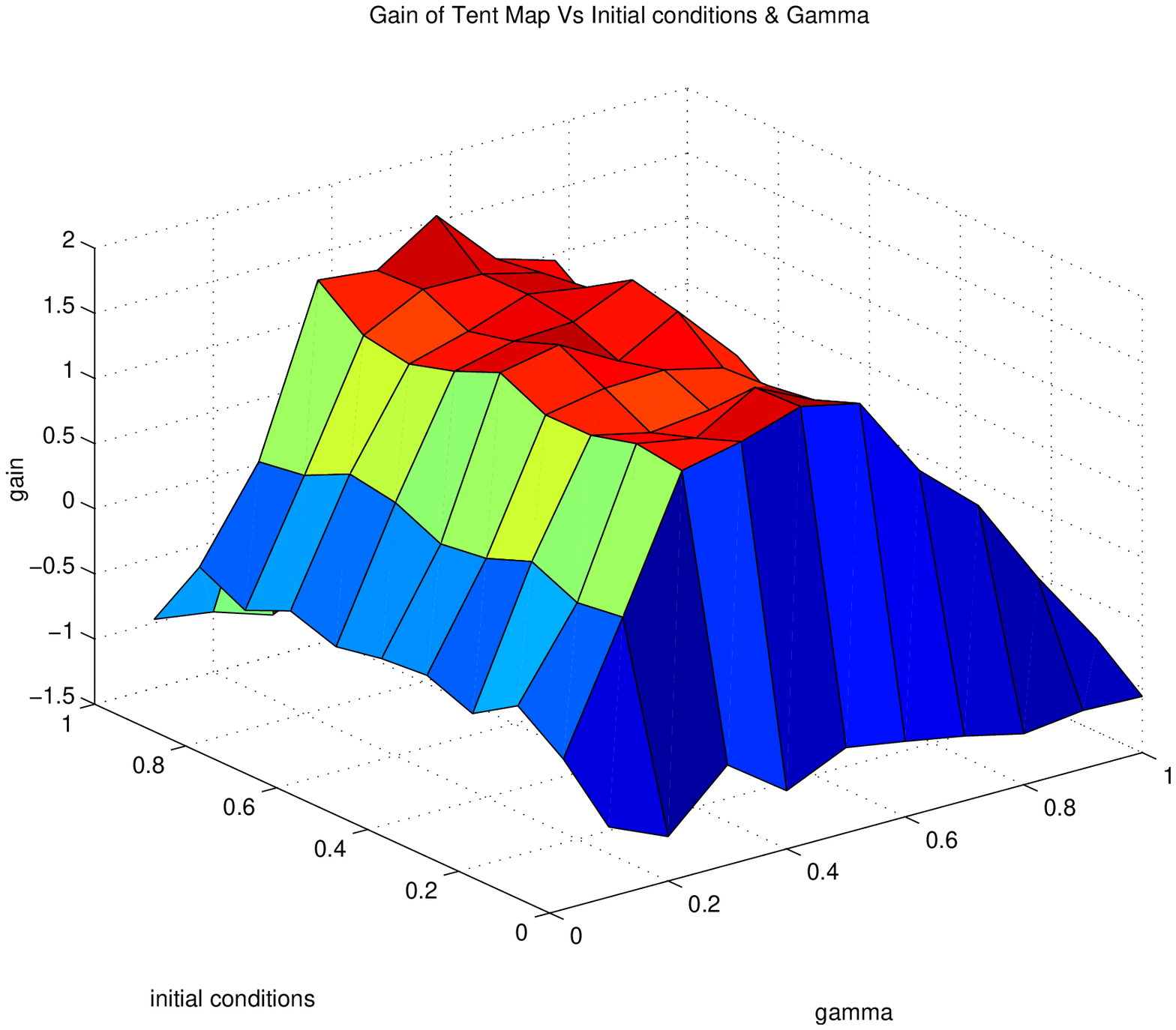}\\
  \label{fig:three:tent}}
\hfill \subfigure[Gain of Gaussian Map switching after 100
games.]{
  \includegraphics[width=6cm]{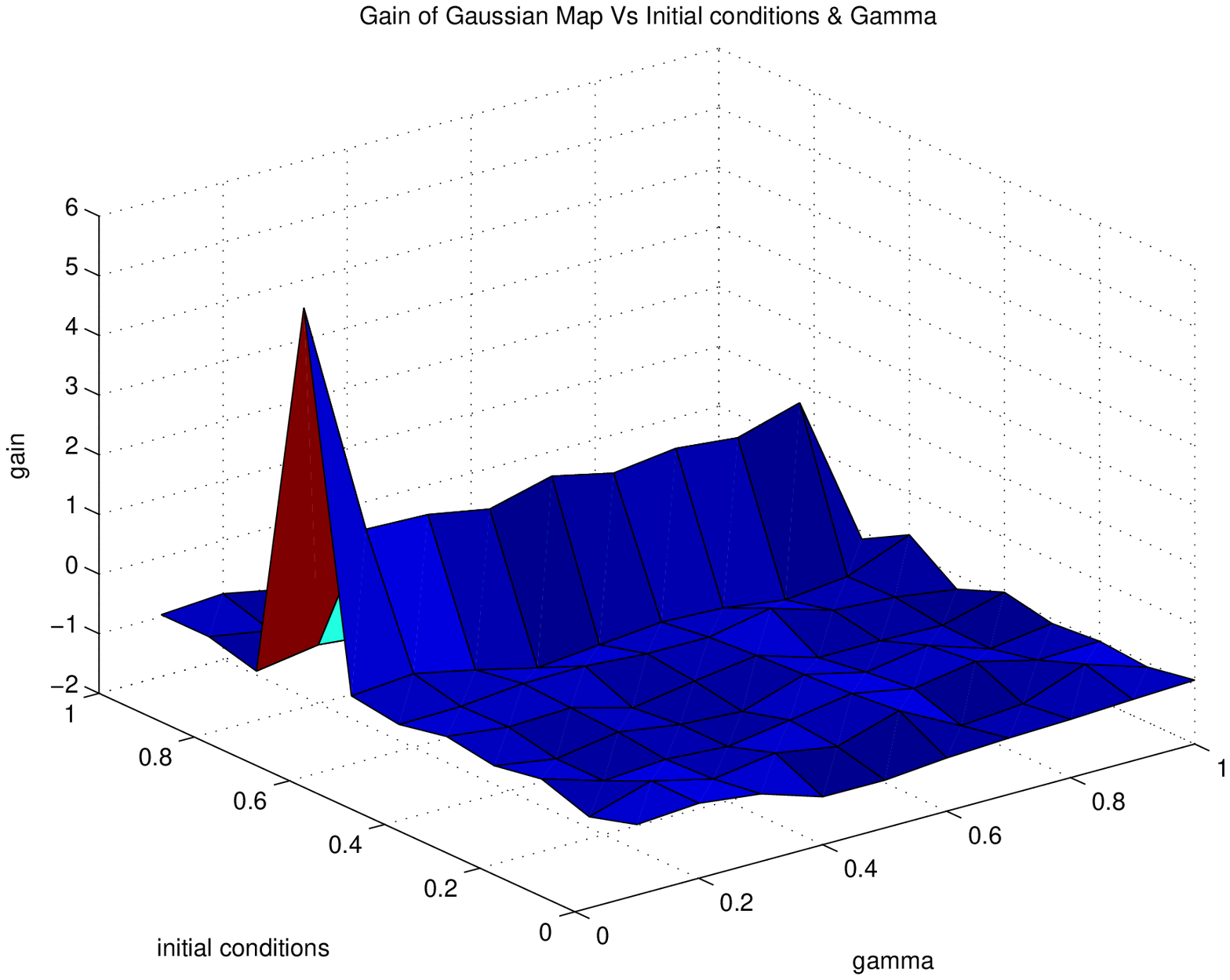}\\
  \label{fig:three:gaus}}
  \hfill \subfigure[Gain of Henon Map switching after 100 games ($a=1.7$, $b=0$).]{
  \includegraphics[width=6cm]{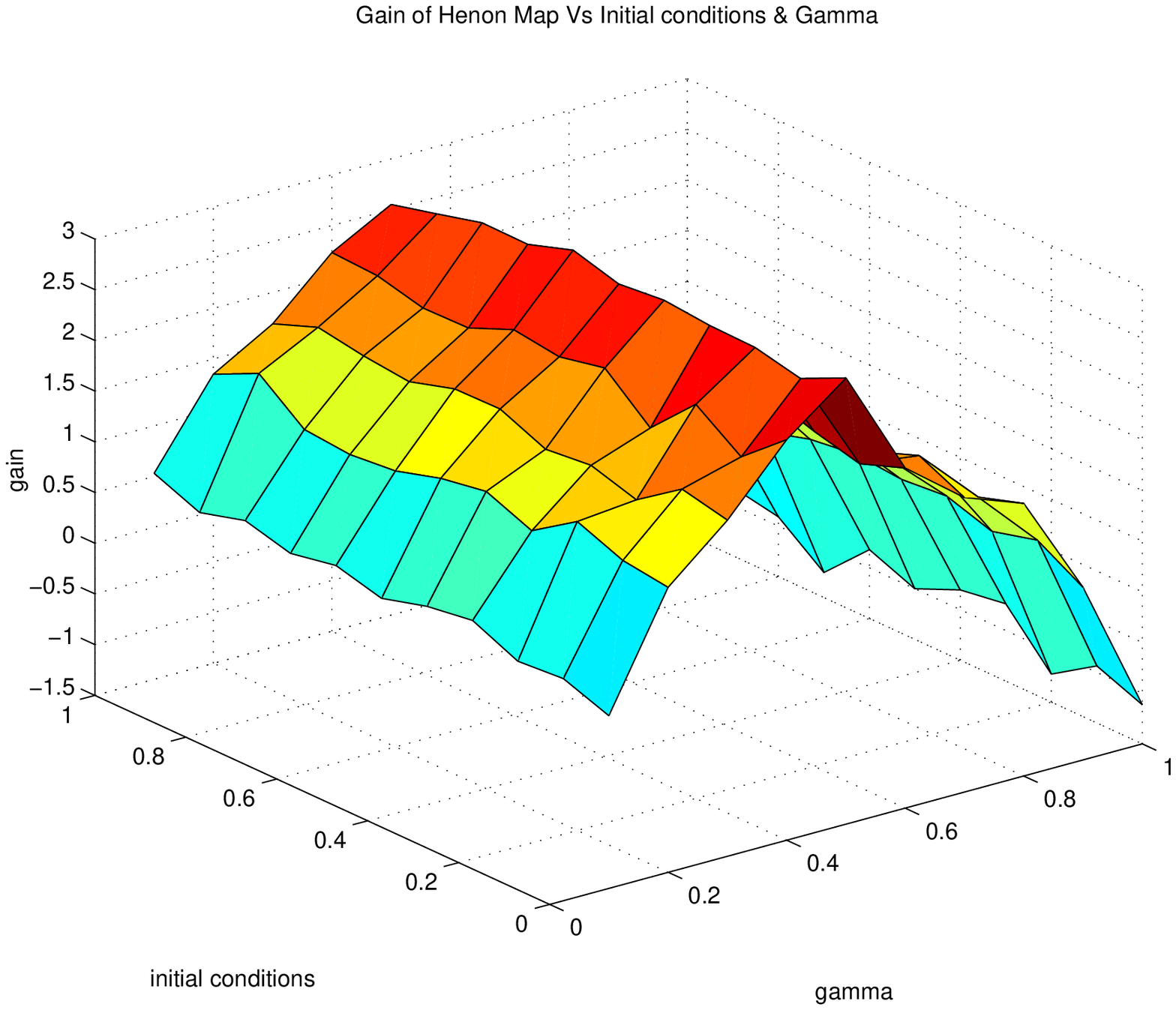}\\
  \label{fig:three:henon}}
    \hfill \subfigure[Gain of Lozi Map switching after 100 games ($a=1.7$, $b=0$).]{
  \includegraphics[width=6cm]{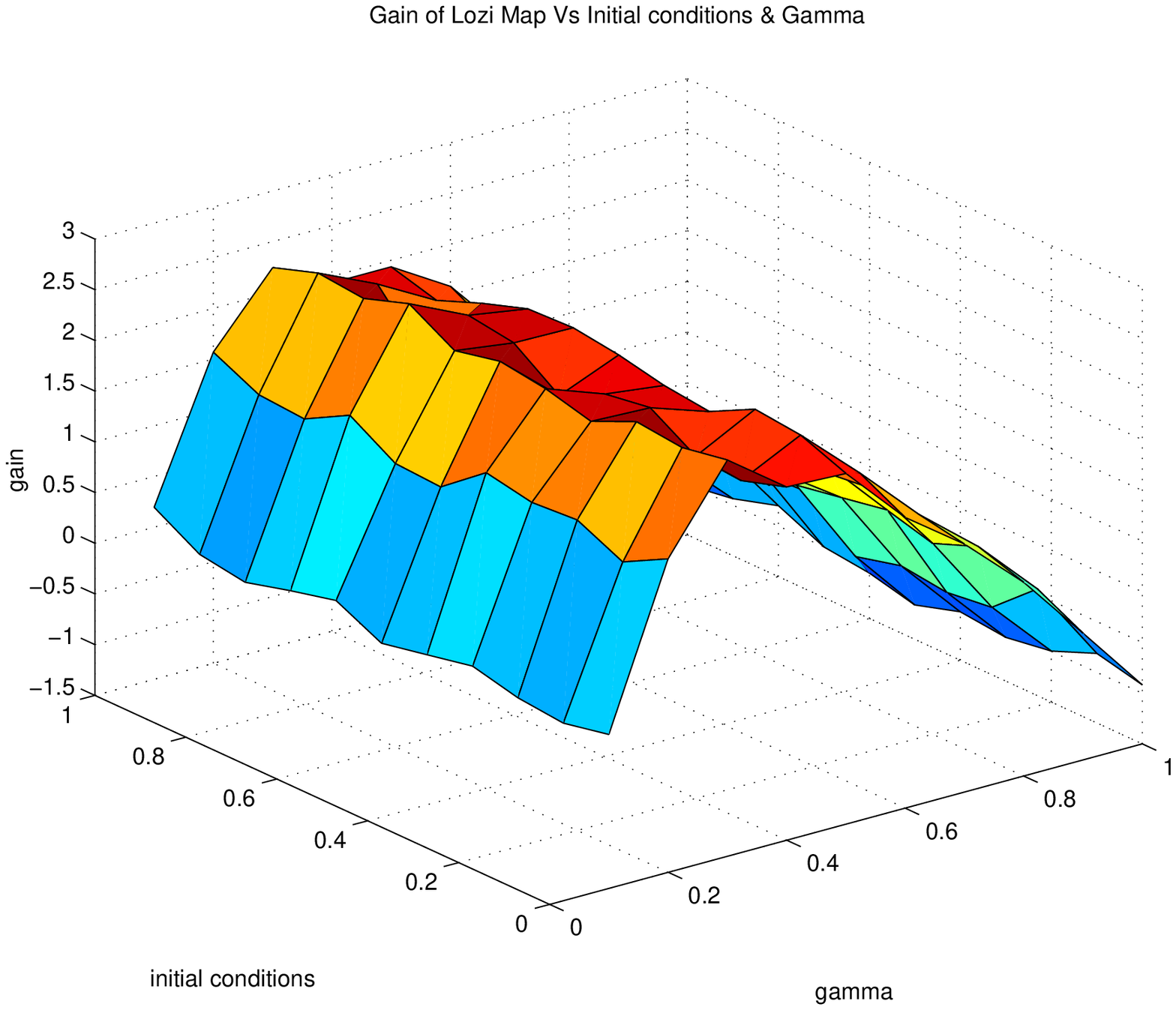}\\
  \label{fig:three:lozi}}
  \caption{Gain of Parrondo's games with chaotic switching against
initial conditions and $\gamma$ value.}\label{fig:three}
\end{figure}

Different combinations of the initial conditions and $\gamma$
values give different rates of winning for Parrondo's game. For
some initial conditions, chaotic switching causes the games to
lose. This occurs when the initial conditions drive the chaotic
sequence towards its attractors as discussed in Section
\ref{sect:initial}. This situation can be explained as playing
Game A or Game B individually since the sequence stays in a
constant value. However, the other initial conditions give the
same rate of winning for a given particular value of $\gamma$ and
coefficient(s) of the chaotic generator. Since the $\gamma$ value
decides the proportion of games played, $\gamma$ value can make
the games either win or lose, as long as the initial condition is
not in the losing region. To obtain optimized or maximum rate of
winning, the capital of the games, after 100 games averaged over
5,000 trials against initial conditions and $\gamma$ value is
plotted. These 3-dimensional diagrams are shown in
Figure~\ref{fig:three:logis} to Figure~\ref{fig:three:gaus}.

For Henon and Lozi Maps, the 3-dimensional diagrams of gain after
100 games against initial conditions and $\gamma$ are plotted
using the simplified maps explained in Section \ref{sect:win:vars}
in order to obtain maximum rate of winning. They are shown in
Figure \ref{fig:three:henon} and Figure \ref{fig:three:lozi}.

When $\gamma$ and the initial condition are set to 0.5 and 0.1
respectively for a Logistic sequence with $a=3.74$, the maximum
gain averaged over 50,000 trials, is found to have value of 6.2
after 100 games.

\section{Effect of initial conditions and $\gamma$ value on the
proportion of Game A played}

The initial conditions of all the chaotic generators have no
affect on the proportion of Game A played. However, the proportion
of Game A played is significantly dependent on $\gamma$ value.
This is because the $\gamma$ value is acting as a threshold value
on deciding whether the next game played should be Game A or B.
The $\gamma$ value draws the boundary of regions between Game A
and B in the normalized time series plot of the chaotic sequence.

\section{Comparing different switchings under same proportion
of Game A played} 
\begin{figure}[p]\begin{center}
  \includegraphics[width=10cm]{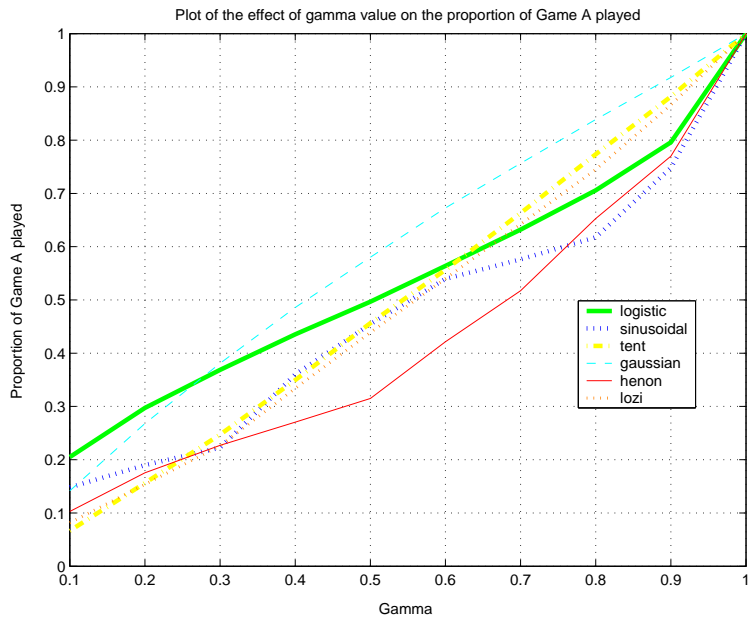}\\
  \caption{Plot of effect of $\gamma$ on the proportion of Game A played.}\label{fig:propA}\end{center}
\end{figure}

\begin{figure}[p]\begin{center}
  \includegraphics[width=11cm]{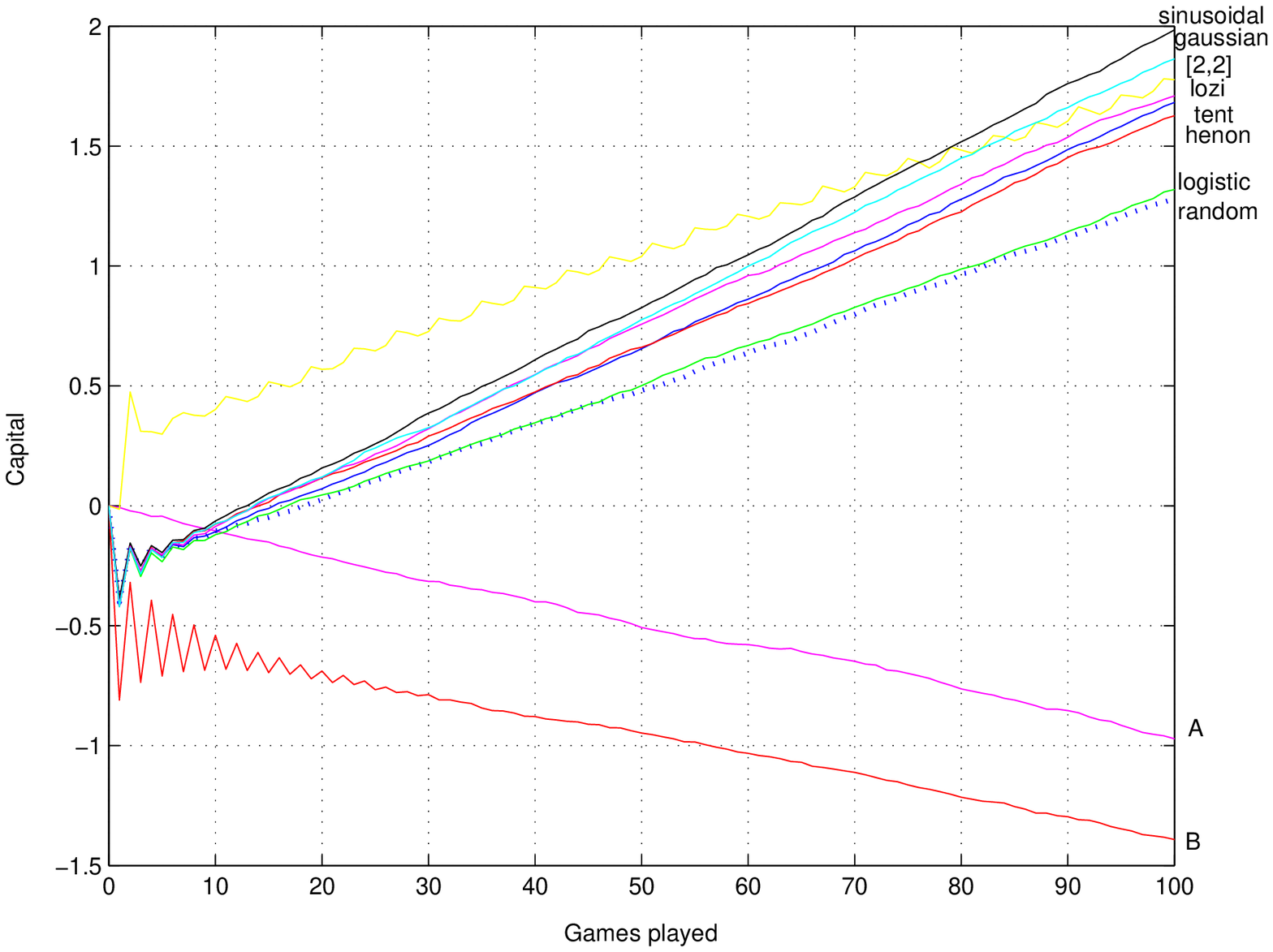}\\
  \caption{Capital under different switching regimes for 100 games (averaged over 50,000 trials).}\label{fig:game:chaos}\end{center}
\end{figure}

The performance of different switchings is based
on the rate of winning Parrondo's games. The higher the rate of
winning, the better the performance is. To compare the performance
of the switchings in a fair manner, a normalization procedure has
to be properly carried out. One suggested way is to compare
switchings with the same proportion of Game A and B played. Since
the proportion of Game A and B played depends on $\gamma$,
$\gamma$ is used to adjust the proportion of Game A played to a
certain fixed value (say 0.5) for all the switchings. The graph
showing the relationship of proportion of Game A played and
$\gamma$ value for chaotic switchings is plotted in Figure
\ref{fig:propA}. The chosen fixed proportion of Game A played for
all the chaotic switchings is 0.5, since the proportion of Game A
played for periodic sequence of [AABB...] or [2,2], is 0.5. Hence,
$\gamma$ is used to obtain 0.5 proportion of Game A played for all
the chaotic switchings and random switching. The $\gamma$ value
that corresponds to 0.5 proportion of Game A played for respective
chaotic switchings can be found in Figure \ref{fig:propA} and
Table \ref{tab:figures}. Under this condition, the rates of
winning of the games with different switchings can be properly
compared. The simulation results of all chaotic switchings
discussed together with random and periodic [2,2] switchings are
plotted in Figure \ref{fig:game:chaos}. This shows that Parrondo's
games with chaotic switching can give higher rate of winning
compared to one with random switching but may or may not be higher
than one with periodic switching.\footnote{It is hard to compare
chaotic switching and periodic switching in a fair manner. This is
because a chaotic switching contains periodic behavior and there
are an infinite number of ways of constructing a periodic
switching} It is found that a particular chaotic switching gives
its highest rate of winning when its sequence is periodic with a
short period.
\section{Conclusion}
The proportion of Game A and B played must be equal for all
switching strategies in order to compare Parrondo's games in a
fair manner. Parrondo's games with chaotic switching can give
higher rate of winning compared to random switching. The rate of
winning obtained from chaotic switching is controlled by the
coefficient(s) defining the chaotic generator, initial conditions
and proportion of Game A played. When a chaotic switching
approaches periodic behavior with a short period, it gives the
highest rate of winning on Parrondo's games. From simulation
results, combination of Game A and B in the pattern
[ABABB...]\footnote{This periodic sequence is found to occur when
$a=3.74$, $\gamma=0.5$, and initial condition = 0.1 in the
Logistic Map averaged over 50,000 trials. It gives gain of 6.2
after 100 games} is found to give the highest rate of winning.

\acknowledgments     
Funding from GTECH Corp. is gratefully acknowledged.
\bibliography{chaosreport}   
\bibliographystyle{spiebib}   

\begin{table}[ht]
\caption{Table showing all the parameter values used in all figures.} \label{tab:figures}
\begin{center}
\begin{tabular}{|c|c|c|c|c|c|} 
\hline \textbf{Figure} & \textbf{\emph{a}} & \textbf{\emph{b}} &
\textbf{$\gamma$ }& \textbf{Initial condition} & \textbf{No. of
trials/}\\

&  & & &  & \textbf{samples}\\
 \hline
\ref{fig:time:logis} & 4&- &- &0.1&100 samples\\ 
\hline
\ref{fig:time:random} & -&- &- &-&100 samples \\
\hline
\ref{fig:phase:logis} & 4&-& -&0.1&5,000 samples\\ 
\hline
\ref{fig:phase:random} & -& -& -& -&5,000 samples\\
\hline
\ref{fig:bifur:logis}& 0 to 4& -& -& 0.1&250 samples\\
\hline
\ref{fig:bifur:sinus} & 2.0 to 2.4& -& -& 0.5&250 samples\\
\hline
\ref{fig:bifur:tent} & 0 to 2& -& -& 0.1&250 samples\\
\hline
\ref{fig:time:logis374} & 3.74&- &- &0.1 &100 samples\\
\hline
\ref{fig:dist:logis374} & 3.74&- &- &0.1 &50,000 samples\\
\hline
\ref{fig:dist:logis} & 4&- &- &0.1 &50,000 samples\\
\hline
\ref{fig:threeAB:henonAB} & 0 to 4 step 0.1& 0 to 4 step 0.1& 0.5&[x,y]=[0,0]&5,000 trials\\
\hline
\ref{fig:threeAB:loziAB} & 0 to 4 step 0.1& 0 to 4 step 0.1& 0.5 &[x,y]=[0,0]&5,000 trials\\
\hline
\ref{fig:three:logis} & 4& -& 0 to 1 step 0.1&0 to 1 step 0.01 &5,000 trials\\
\hline
\ref{fig:three:sinus} & 2.27& -&0 to 1 step 0.1 & 0 to 1 step 0.1&5,000 trials\\
\hline
\ref{fig:three:tent}&1.9 &- &0 to 1 step 0.1 & 0 to 1 step 0.1&5,000 trials\\
\hline
\ref{fig:three:gaus} &- &- &0 to 1 step 0.1 & 0 to 1 step 0.1&5,000 trials\\
\hline
\ref{fig:three:henon} & 1.7& 0& 0 to 1 step 0.1&0 to 1 step 0.1 &5,000 trials\\
\hline
\ref{fig:three:lozi}& 1.7& 0&0 to 1 step 0.1 & 0 to 1 step 0.1&5,000 trials\\
\hline
\ref{fig:propA} & same as below& same as below&0 to 1 step 0.1 & same as below&5,000 trials\\
\hline
\ref{fig:game:chaos} &- &- &- & -&-\\
Logistic &4 &- &0.50 & 0.1&50,000 trials\\
Sinusoidal &2.27 &- &0.55 & 0.5&50,000 trials\\
Tent &1.9 &- & 0.55 &  0.8&50,000 trials\\
Gaussian&- &- &0.41 & 0.701&50,000 trials\\
Henon &1.7 &0 &0.68 & [x,y]=[0,0]&50,000 trials\\
Lozi &1.7 &0 & 0.55 & [x,y]=[0,0]&50,000 trials\\

\hline
\end{tabular}
\end{center}
\end{table}

\end{document}